\documentclass{ws-procs975x65}
\usepackage[utf8]{inputenc}
\usepackage[T1]{fontenc}

\begin{document}

\title{AdS-MAXWELL BF THEORY AS A MODEL OF GRAVITY AND BI-GRAVITY}

\author{A. Borowiec$^*$, J. Kowalski-Glikman$^{**}$ and M. Szcz\k{a}chor$^{***}$}

\address{ Institute for Theoretical Physics, University of Wrocław,\\
Wroclaw, 50-204, Poland\\
$^{*}$E-mail: borow@ift.uni.wroc.pl\\
$^{**}$E-mail: jkowalskiglikman@ift.uni.wroc.pl\\
$^{***}$E-mail: misza@ift.uni.wroc.pl}

\begin{abstract}
This article presents an extended model of gravity obtained by
gauging the AdS-Mawell algebra. It involves additional fields that
 shift the spin connection, leading effectively to theory of two
 independent
connections. Extension of algebraic structure by another tetrad
gives rise to the model described by a pair of Einstein equations.
\end{abstract}


\bodymatter
\newcommand{\so}{\mathfrak{so}}
\newcommand{\ISO}{{\rm ISO}}
\newcommand{\iso}{\mathfrak{iso}}
\def\MM{MacDowell--Mansouri }
\def\bp{\partial\!\!\!/}
\def\bop[#1]{\overleftrightarrow{\partial}_{#1}}
\def\be{\begin{equation}}
\def\ee{\end{equation}}
\def\beq{\begin{eqnarray}}
\def\eeq{\end{eqnarray}}
\def\bn{\begin{eqnarray*}}
\def\en{\end{eqnarray*}}
\def\slas{\!\!\!/}
\def\P{\Phi}
\def\p{\phi}
\def\w{\omega}
\def\W{\Omega}
\def\O{{\cal{O}}}
\def\a{\alpha}
\def\b{\beta}
\def\s{\sigma}
\def\S{\Sigma}
\def\d{\delta}
\def\D{\Delta}
\def\g{\gamma}
\def\t{\theta}
\def\T{\Theta}
\def\G{\Gamma}
\def\z{\zeta}
\def\k{\kappa}
\def\Z{\Psi}
\def\si{\psi}
\def\pd{\partial}
\def\e{\epsilon}
\def\n{\nu}
\def\m{\mu}
\def\r{\rho}
\def\t{\theta}
\def\ta{\tau}
\def\R{\Rho}
\def\l{\lambda}
\def\L{\Lambda}
\def\cL{{\cal{L}}}
\def\cH{{\cal{H}}}
\def\cP{{\cal{P}}}
\def\cU{{\cal{U}}}
\def\cB{{\cal{B}}}
\def\cN{{\cal{N}}}
\def\cT{{\cal{T}}}
\def\cC{{\cal{C}}}
\def\cD{{\cal{D}}}
\def\cF{{\cal{F}}}
\def\cJ{{\cal{J}}}
\def\cK{{\cal{K}}}
\def\cM{{\cal{M}}}
\def\cZ{{\cal{Z}}}
\def\cR{{\cal{R}}}
\def\sl{\not\!}
\def\slD{\not\!D}
\def\la{\langle}
\def\ra{\rangle}
 $\ \ $\\
It was shown in the recent works\cite{Gomis:2009dm, Bonanos:2010fw}
that it is possible to extend the Poincare algebra by extra
charges, which are related to a constant background Maxwell
field\cite{Schrader:1972zd}. Such algebra is called Maxwell algebra.
Unfortunately Maxwell algebra can not be used to gauging the theory in
geometrical way\cite{MacDowell:1977jt}, because, like the
original Poincare algebra, it does not possess a non-degenerate
ad-invariant inner product. Therefore as a first step to gauge
Maxwell algebra one has to form its (Anti) de Sitter extension,
which reads\cite{Soroka:2011tc, Soroka:2006aj,Durka:2011nf} \begin{eqnarray}
\nonumber&&[\mathcal{M}_{ab},\mathcal{M}_{cd}]=-i(\eta_{ac}\mathcal{M}_{bd}+\eta_{bd}\mathcal{M}_{ac}-\eta_{ad}\mathcal{M}_{bc}-\eta_{bc}\mathcal{M}_{ad})\,,\\
\nonumber &&[\mathcal{M}_{ab},\mathcal{P}_c]=-i(\eta_{ac}\mathcal{P}_{b}-\eta_{bc}\mathcal{P}_{a})\,, \\
\label{jeden} &&[\mathcal{P}_{a},\mathcal{P}_{b}]=i(\mathcal{M}_{ab}+k\mathcal{Z}_{ab}),\qquad [\mathcal{Z}_{ab},\mathcal{P}_{c}]=0\\
\nonumber &&[\mathcal{M}_{ab},\mathcal{Z}_{cd}]=-i(\eta_{ac}\mathcal{Z}_{bd}+\eta_{bd}\mathcal{Z}_{ac}-\eta_{ad}\mathcal{Z}_{bc}-\eta_{bc}\mathcal{Z}_{ad}), \\
\nonumber &&[\mathcal{Z}_{ab},\mathcal{Z}_{cd}]=+ik(\eta_{ac}\mathcal{Z}_{bd}+\eta_{bd}\mathcal{Z}_{ac}-\eta_{ad}\mathcal{Z}_{bc}-\eta_{bc}\mathcal{Z}_{ad})\,.
\end{eqnarray} with $k=+1$ for dS-Maxwell, $k=-1$ for AdS-Maxwell algebra and \\ $a,b,\dots=0,\dots,3$.

Gauging this algebra one gets the connection
\begin{equation}
\mathbb{A}_{\mu}=\frac{1}{2}\omega_\mu{}^{ab} \mathcal{M}_{ab}+\frac{1}{\ell}e_\mu^{a} \mathcal{P}_{a}+\frac{1}{2}h_\mu^{ab} \mathcal{Z}_{ab}\,,
\end{equation}
and its curvature
\begin{equation}
\mathbb{F}_{\mu\nu}=\frac12\, F_{\mu\nu}^{ab}\, \mathcal{M}_{ab} +\frac{1}{\ell}T_{\mu\nu}^{a}\,\mathcal{P}_{a}+ \frac12\, G_{\mu\nu}^{ab}\, \mathcal{Z}_{ab}
\end{equation}
which makes it possible to construct a gauge invariant action in the
form of a constrained topological BF theory\cite{Freidel:2005ak}.
This action reads\cite{Durka:2011nf}
\begin{eqnarray}\label{444}
 16\pi\, S(A,B)&=& \int 2(B^{a4}\wedge F_{a4}-\frac{\beta}{2}B^{a4}\wedge B_{a4}) \nonumber\\
& &  +B^{ab}\wedge F_{ab}- \frac{\beta}{2}B^{ab}\wedge B_{ab}- \frac{\alpha}{4}\epsilon^{abcd} B_{ab}\wedge B_{cd}\nonumber\\
& &+C^{ab}\wedge G_{ab}- \frac{\rho}{2}C^{ab}\wedge C_{ab} - \frac{\sigma}{4}\epsilon^{abcd} C_{ab}\wedge C_{cd}\nonumber\\
& & - \beta  C^{ab}\wedge B_{ab}- \frac{\alpha }{2}\epsilon^{abcd} C_{ab}\wedge B_{cd}\label{full_l}\,,
\end{eqnarray}
where $C^{ab}, B^{ab}, B^{a4}$ are auxiliary
fields. By construction the action (\ref{full_l}) is manifestly
diffeomorphism-invariant and possess local Lorentz and Maxwell
symmetries, but the translational part of (A)dS-Maxwell symmetry,
generated by $P$ is broken\cite{MacDowell:1977jt,Freidel:2005ak}.
After eliminating the auxiliary fields by solving their field
equations the action (\ref{444}) takes form
\begin{eqnarray}
 \label{dwa} 16 \pi S(\omega,h,e)&=& \int\left( \frac{1}{4}M^{abcd} F_{ab}\wedge F_{cd}-\frac{1}{\beta\ell^2} \,T^a \wedge T_a\right)\\
 & &\quad+{\int\frac{1}{4}N^{abcd} (G_{ab}+F_{ab})\wedge (G_{ab}+F_{ab})}\,,\nonumber
\end{eqnarray}
where
$$
M^{ab}{}_{cd} = \frac\alpha{\alpha^2+\beta^2}\,
\left(\frac\beta\alpha\, \delta^{ab}_{cd}-
\epsilon^{ab}{}_{cd}\right)
$$
and
$$   N^{abcd}=\frac{(\sigma-\alpha)}{(\sigma-\alpha)^2+(\rho-\beta)^2}\left(\frac{\rho-\beta}{\sigma-\alpha}\delta^{abcd} -\epsilon^{abcd} \right)$$
Since the last term in (\ref{dwa}) is a topological
invariant\footnote{It is
 combination of the Euler and Pontryagin invariant for shifted connection $\omega+h$.}  this action describe pure gravity.

The algebra (\ref{jeden}) can be further extended by adding yet
another translational generator $\mathcal{R}_a$ with the commutation
relations
\begin{equation}
[\mathcal{R}_{a},\mathcal{R}_{b}]=i\mathcal{Z}_{ab},\qquad [\mathcal{P}_{a},\mathcal{R}_{c}]=0\, ,
\end{equation}
\be
[\mathcal{M}_{ab},\mathcal{R}_c]=-i(\eta_{ac}\mathcal{R}_{b}-\eta_{bc}\mathcal{R}_{a}), \quad [\mathcal{Z}_{ab},\mathcal{R}_c]=-i(\eta_{ac}\mathcal{R}_{b}-\eta_{bc}\mathcal{R}_{a})\,.
\ee  The gauge connection becomes
\begin{equation}\label{sylwester}
\mathbb{A}_{\mu}=\frac{1}{2}\omega_\mu{}^{ab} \mathcal{M}_{ab}+\frac{1}{\ell}e_\mu^{a} \mathcal{P}_{a}+\frac{1}{2}h_\mu^{ab} \mathcal{Z}_{ab}+\frac{1}{\ell'}f_\mu^a \mathcal{R}_a\,,
\end{equation}
and the gauge curvatures take the form
 \begin{eqnarray}
 F_{\mu\nu}^{ab}&=&R^{ab}_{\mu\nu}+\frac{1}{\ell^2}( e^a_\mu e^b_\nu- e^a_\nu e^b_\mu), \quad T_{\mu\nu}^{a}=D^\omega_\mu e^a_\nu -D^\omega_\nu e^a_\mu\,,\label{5b}\\
 G_{\mu\nu}^{ab}&=&D^\omega_\mu h^{ab}_\nu -D^\omega_\nu
 h^{ab}_\mu\nonumber-\frac{1}{\ell^2}( e^a_\mu e^b_\nu- e^a_\nu
 e^b_\mu)\\
&+&(h^{ac}_\mu h^{\quad b}_{\nu\,c}-h^{ac}_\nu h^{\quad b}_{\mu\,c}){-\frac{1}{\ell^2}( f^a_\mu f^b_\nu- f^a_\nu f^b_\mu)} \,, \label{5d}\\
Y_{\mu\nu}^a &=& D^\omega_\mu f^{a}_\nu -D^\omega_\nu f_\mu^a+h_{\mu}^{ab}f_{\nu b}-h_{\nu}^{ab}f_{\mu b} .\label{5c}
 \end{eqnarray}
The total action is slightly different from (\ref{full_l}) and
equals
\begin{equation}\label{six}
S=S^{(AdS-Maxwell)} + 2\, C^a \wedge Y_a\,.
\end{equation}
The crucial observation is that after adding the new generator, the action
(\ref{dwa}) is not topological anymore because the expression
(\ref{5d}) has changed. If the connection is shifted by the Maxwell field i.e., $\varpi=\omega+h$,
which is equivalent to changing the basis of the Lie algebra by
$\mathcal{M}\rightarrow \mathcal{M}-\mathcal{Z}$, the last term of (\ref{dwa}) becomes another Einstein
action for connection $\varpi$, tetrad $f$, and with cosmological
constant $\Lambda'$
\begin{equation}
S_{E_2} =\frac1{64\pi G}\,
\int\epsilon_{abcd}\left(H_{\mu\nu}[\varpi]{}^{ab}f_{\rho}^{c}f_{\sigma}^{d}
-\frac{\Lambda'}{3}f_{\mu}^{a}f_{\nu}^{b}f_{\rho}^{ c}f_{\sigma}^{
d}\right)\epsilon^{\mu\nu\rho\sigma}\label{18}\,,
\end{equation}
where $H_{\mu\nu}[\varpi]{}^{ab}$ is curvature of the shifted connection, and $\Lambda'=3/{\ell'}$ (see eq. \ref{sylwester}).
It follows from the field equation derived from (\ref{six}) that the
torsion of the connection $\varpi$ vanishes
\begin{equation}\label{17}
    df^a + \varpi^a{}_b\wedge f^b=\tilde{T}^a=0\,.
\end{equation}
It is worth recalling that the model of Einstein pair was previously
discussed in a similar context in Ref. \refcite{Durka:2011va}, but there the
existence of another tetrad, and the torsion equation was introduced
by hand. Here the tetrad emerges naturally as a result of the
algebra enlargement, and satisfies field equations, which forces
torsion to vanish. However, there is not interaction term between
tetrads, which is crucial in bi-gravity theories such as the $f-g$
model\cite{Isham:1974rh}. It was pointed out in Ref. \refcite{Durka:2011va}
that one can add such gauge invariant term,  but it is not possible
to obtain it by enlarging the (A)dS algebras. We will present  a
more detailed discussion  of these  algebras and their physical
applications 
 in a separate paper.

 \section*{Acknowledgment}
This work has been supported by the grant 2011/01/B/ST2/03354 (AB, JKG); by funds provided by the National Science Center under the agreement
DEC- 2011/02/A/ST2/00294 (JKG); and by the grant 2011/01/N/ST2/00415 (MS).

\bibliographystyle{ws-procs975x65}

\end{document}